\documentclass[useAMS,usenatbib]{mn2e}

\usepackage{graphicx}

\title[$\gamma^2$~Vel: Orbital Solution and Fundamental Parameters]
      {$\gamma^2$ Velorum: Orbital Solution and Fundamental
    		    Parameter Determination with SUSI}
      
\author[North et. al.]
       {J. R. North\thanks{E-mail: j.north@physics.usyd.edu.au},  
	P. G. Tuthill, W. J. Tango and J. Davis \\
	School of Physics, University of Sydney, NSW 2006, Australia}

\begin{document}

\date{Accepted ; Received ; in original form }

\pagerange{\pageref{firstpage}--\pageref{lastpage}} \pubyear{2007}

\maketitle

\label{firstpage}

\begin{abstract}
The first complete orbital solution for the double-lined spectroscopic 
binary system $\gamma^2$~Velorum, obtained from measurements with the 
Sydney University Stellar Interferometer (SUSI), is presented. This system 
contains the closest example of a Wolf-Rayet star and the promise of full 
characterisation of the basic properties of this exotic high-mass system 
has subjected it to intense study as an archetype for its class.
In combination with the latest radial-velocity results, our orbital solution 
produces a distance of $336^{+8}_{-7}$\,pc, significantly more distant
than the {\em Hipparcos} estimation (\citealt{Schaerer97}; 
\citealt{Hucht97}). The ability to fully specify the orbital parameters
has enabled us to significantly reduce uncertainties and our result is
consistent with the VLTI observational point \citep{Millour06}, but
not with their derived distance. Our new distance, which is an order of 
magnitude more precise than prior work, demands critical reassessment of all 
distance-dependent fundamental parameters of this important system. In 
particular, membership of the Vela OB2 association has been reestablished, 
and the age and distance are also in good accord with the population of 
young stars reported by \citet{Pozzo00}.
We determine the O-star primary component parameters to be 
M$_{V}$(O) = $-5.63\pm0.10$\,mag, R(O) = $17\pm2$\,R$_{\sun}$ and 
${\cal M}$(O) = $28.5\pm1.1$\,M$_{\sun}$. These values are consistent with 
calibrations found in the literature if a luminosity class of II--III is 
adopted. The parameters of the Wolf-Rayet component are 
M$_{v}$(WR) = $-4.33\pm0.17$\,mag and 
${\cal M}$(WR) = $9.0\pm0.6$\,M$_{\sun}$. 
\end{abstract}


%
%

\section{Introduction}

The high-luminosity Wolf-Rayet (WR) stars are characterised by
bright emission line spectra produced in hot, high-speed stellar winds
\citep{Hucht92}. Thought to embody the final stable stage in the evolution of 
very massive stars, they are candidate precursors to Type Ib/Ic supernovae 
\citep{Hucht01}. The fast ($v_{\infty} \simeq 10^3$kms$^{-1}$) stellar 
winds of WR stars are responsible for prodigious mass loss 
($10^{-5}$\,M$_{\sun}$\,yr$^{-1}$), eventually stripping the star of its outer 
layers and forming their characteristic He-rich emission-line spectrum. 
Despite the ephemeral nature of this phase, these stars are important not 
only for understanding the evolution of massive stars but also as tracers of 
galactic structure and star formation \citep{Hucht01}.

The double-lined spectroscopic binary $\gamma^2$~Velorum (HR 3207, HD 68273, 
WR 11\footnote{WR number refers to the catalogue of \citet{Hucht01}.}) 
contains the brightest example of the WR type in the sky
and has a combined visual magnitude $V \simeq 1.8$ with an O-star companion. 
It has a {\em Hipparcos} determined distance of $258^{+41}_{-31}$\,pc
(\citealt*{Schaerer97}; \citealt{Hucht97}), placing $\gamma^2$~Vel closer to 
Earth than any other WR star by a factor of approximately two 
(\citealt{Morris00}; \citealt{Setia01}). Its proximity is not the only 
reason that it is a key target for astronomers: the orbital motion of the 
two stars makes $\gamma^2$~Vel one of the few systems for which a direct 
mass determination of the components can be achieved. As such, $\gamma^2$ 
Vel provides critical tests for the theory of WR and massive star 
evolution. It is therefore not surprising that numerous studies of this 
system have been made, but in spite of its seemingly textbook status, 
considerable uncertainties remain. 

Spectroscopic analysis of $\gamma^2$~Vel has been hampered by 
entanglement of the O-star and WR wind emission: all absorption features of 
the brighter O-star component are blended with emission lines from the WR 
wind \citep{DeMarco00}. Accounting for the WR emission is not trivial 
because {\em a)} multiple emission lines may contribute to the blend, 
{\em b)} the shape of the emission line may differ significantly from a 
Gaussian, {\em c)} the emission lines vary in time with some excess emission 
attributed to wind-wind collision phenomena, {\em d)} the extended wings of 
the strong O-star absorption are lost in the blend 
(\citealt{Schmutz97}; \citealt{DeMarco99}; \citealt*{Eversberg99}).
This has resulted in numerous luminosity and spectral class identifications
of the O-star in the literature: O7.5 \citep{Ganesh67}, O8 \citep{Baschek71},
O9~I \citep{Conti72}, O8~III \citep{Schaerer97} and O7.5~III--V 
\citep{Hucht01}.The WR component has retained its original classification of 
type WC8 by \citet{Smith68a}.

The first to determine the period and some of the orbital elements by
radial velocity measurements were \citet{Ganesh67}. Further 
spectroscopic studies, for example \citet{Niemela80}, \citet*{Pike83}
and \citet{Moffat86}, continued to add to our knowledge of the system
but discrepancies could not be resolved and observations from the {\em IUE}
could {\em `only  be seen as muddying some already murky waters'} 
\citep{Stickland90}. By correcting the O-star absorption lines due to the WR 
emission, \citet{Schmutz97} believe they have the most definitive spectroscopic 
orbit to date. However, this type of analysis cannot determine all of the 
orbital elements. For example, the inclination has been estimated from the 
polarimetric analysis of \citet{StLouis87} and further constraints were
imposed by \citet{Schmutz97}, \citet{DeMarco99} and \citet{DeMarco00} who
used the {\em Hipparcos} distance, stellar models and spectroscopy.

Due to the high angular resolution needed, only interferometric techniques 
can currently be employed to accurately constrain the remaining orbital 
parameters. However, the southerly declination ($-47\degr20\arcmin$) of
$\gamma^2$~Vel places it out of reach for many interferometers of 
sufficient resolving power. \citet{HBrown70} used the Narrabri Stellar 
Intensity Interferometer (NSII) to study this system, determining 
the angular semi-major axis to be $4.3 \pm 0.5$\,mas. Unfortunately, the 
signal-to-noise of their observations necessitated the adoption of the 
parameters of \citet{Ganesh67} which have now been superseded by those of
\citet{Schmutz97}. Furthermore, due to the lack of observations, the 
inclination could not be adequately constrained and a value $i=70\degr$ was 
assumed from theoretical arguments. The Very Large Telescope Interferometer 
(VLTI) has observed $\gamma^2$~Vel and produced a single angular 
separation and position angle \citep{Millour06} but has yet to improve the 
orbital parameters.

Reliance on distance estimations and stellar models has paved the way
for further analysis of fundamental parameters, although this path has been
fraught with controversy. The pre-{\em Hipparcos} distance of $\gamma^2$~Vel 
was inferred from the assumed membership of the Vela OB2 association or from 
statistical absolute magnitudes of similar or nearby stars. \citet{Baschek71} 
assumed equidistance with the visual companion $\gamma^1$~Vel while 
\citet{Conti72} adopted 460\,pc from estimates of nearby stars. The 
interferometric study of \citet{HBrown70} combined with the spectroscopy of 
\citet{Ganesh67} produced the first dynamical parallax of this system yielding 
$350 \pm 50$\,pc, although this distance is affected by the assumed values
of some orbital parameters. In the 1980s, the distance to $\gamma^2$~Vel 
was assumed to be equal to that of the Vela OB2 association (450pc) with 
sufficient certainty as to be used in the determination of galactic WR 
intrinsic parameters \citep{Hucht88}. The publication of the significantly
different {\em Hipparcos} parallax distance of $258^{+41}_{-31}$\,pc by 
\citet{Schaerer97} and \citet{Hucht97} caused a critical reassessment of all
$\gamma^2$~Vel parameters that are distance dependent -- the absolute
magnitude of the components, the spectral class of the O-star primary, the 
mass-loss of the WR star etc. The {\em Hipparcos} distance has been
challenged by the discovery of a low-mass, pre-main-sequence stellar 
association in the direction of $\gamma^2$~Vel and the Vela OB2 association.
\citet{Pozzo00} argue these low-mass stars are associated with $\gamma^2$~Vel
and approximately coeval at a distance consistent with earlier estimates of 
360--490\,pc. The single measurement by the VLTI has yielded a measure of
the distance to $\gamma^2$~Vel of $368^{+38}_{-13}$\,pc (\citealt{Malbet06};
\citealt{Millour06}) placing the system farther from the {\em Hipparcos} 
distance and closer to the Vela OB2 association.

In Section~\ref{fit} we provide the first complete orbital solution for 
$\gamma^2$~Vel determined by long-baseline optical interferometry. In 
combination with the latest radial velocity measurements, the tightly
constrained distance, spatial scale and mass of the components are 
quantified and compared to previous estimates in Section~\ref{disc}. 
The details of our observations and a description of 
the parameter fitting procedure are described in Section~\ref{obs} and 
Section~\ref{fit} respectively.

%
%

\section{Observations and Data Reduction}
\label{obs}

\begin{table}
    \caption{Adopted parameters of reference stars used during observations.}
    \label{cal_table}
    \begin{tabular}{@{}cccccc@{}}
	\hline
    HR 	 & Name 	& Spectral & V    & UD Diameter       & Separation\\
	 &		&   Type   &      &  (mas)    	      & from $\gamma^{2}$ Vel\\
	\hline
    3090 & J Pup   	& B0.5~Ib   & 4.24 & $0.18 \pm 0.03^a$ &  2.96\degr \\
    3117 & $\chi$ Car   & B3I~Vp    & 3.47 & $0.36 \pm 0.06^a$ &  6.02\degr \\
    3165 & $\zeta$ Pup  & O5f	   & 2.25 & $0.41 \pm 0.03^b$ &  7.41\degr \\
    3468 & $\alpha$ Pyx & B1.5~III  & 3.68 & $0.27 \pm 0.06^a$ & 15.54\degr \\
	\hline
    \end{tabular}
    \newline
    \begin{tabular}{@{}ll}
	Notes:  & $^a$ Based on NSII measurements of similar type stars. \\
		& $^b$ \citet*{HBrown74}.\\
    \end{tabular}	
\end{table}

Measurements of the squared visibility (i.e. the squared modulus of the 
normalised complex visibility) or $V^2$ were completed on a total of 24 
nights using the Sydney University Stellar Interferometer (SUSI,
\citealt{Davis99}). Interference fringes were recorded with the red 
beam-combining system using a filter with centre wavelength and full-width 
half-maximum of 700\,nm and 80\,nm respectively. This system was outlined in 
\citet{Tuthill04} and is to be described in greater detail in Davis et al. 
(in preparation).

Interference fringes produced by a pupil-plane beam-combiner were
modulated by repeatedly scanning the optical delay about the white
light fringe position. The two outputs of the beam-combiner were detected 
by avalanche photo-diodes. An observation unit consisted of a set of 1000 
scans traversing 140\,$\mu$m in optical delay digitised into 1024 steps of 
0.2\,ms samples.

During post-processing of the data, the two signals were differenced to 
mitigate noise introduced by scintillation and the squared visibility was 
estimated after bias subtraction for each observation. A nonlinear function 
was applied to partially correct for residual seeing effects
\citep{Ireland06}. 

Reference stars close $\gamma^2$~Vel on the sky, chosen to provide 
calibration of the system response, were interleaved with the target 
observations. The system response as a function of time was quantified 
using the adopted stellar parameters of the calibrator stars given in 
Table~\ref{cal_table}. Measurements of $V^2$ were finally produced by 
linearly interpolating the system response to the time of target observation 
and scaling the observed squared visibility appropriately. This procedure 
resulted in a total of 278 estimations of $V^2$ as summarised in 
Table~\ref{obs_table}. The majority of the observations were completed in 
excellent conditions and the atmospheric correction to the final $V^2$ 
values was small with a mean value of the order 2 percent.

\begin{table*}
  \centering
  \begin{minipage}{135mm}
    \caption{Summary of observational data.  The night of the observation is 
             given in cols 1 and 2 as a calendar date and a mean MJD. Col 3 is 
	     the mean orbital phase calculated from the values in 
	     Table~\ref{fit_table}. The baseline and the mean projected 
	     baseline (in units of metres) are given in Col 4-5 respectively.
	     Reference stars and the the number of squared visibility 
	     measures for a night are listed in the last two columns.} 
    \label{obs_table}
    \begin{tabular}{@{}ccccclc}
	\hline
	Date  	& MJD		&  Phase    & Nominal  & Projected & Reference & \# V$^2$\\
		&		&	    & Baseline & Baseline  &	Stars    &	   \\
	\hline
2005 Mar 10 & 53439.52 & 0.26 & 80 & 75.52	&$\zeta$ Pup, $\chi$ Car, J Pup, $\alpha$ Pyx	& 10 \\	
2005 Mar 11 & 53440.48 & 0.28 & 80 & 75.93	&$\zeta$ Pup, $\chi$ Car, $\alpha$ Pyx	& 15 \\	
2005 Mar 12 & 53441.48 & 0.29 & 80 & 75.78	&$\zeta$ Pup, $\chi$ Car, $\alpha$ Pyx	& 12 \\	

2005 Dec 17 & 53721.70 & 0.86 & 80 & 76.19	&$\zeta$ Pup, $\chi$ Car, $\alpha$ Pyx	&  9  \\		
2005 Dec 18 & 53722.69 & 0.87 & 80 & 76.18	&$\zeta$ Pup, $\chi$ Car, J Pup		& 13 \\
2005 Dec 19 & 53723.68 & 0.88 & 80 & 76.24	&$\zeta$ Pup, $\chi$ Car, J Pup		& 12 \\
2005 Dec 20 & 53724.68 & 0.89 & 80 & 76.19	&$\zeta$ Pup, $\chi$ Car, J Pup		& 12 \\
2005 Dec 23 & 53727.62 & 0.93 & 80 & 76.02	&$\zeta$ Pup, $\chi$ Car			&  6 \\
2005 Dec 29 & 53733.67 & 0.01 & 80 & 76.02	&$\zeta$ Pup, $\chi$ Car			& 14 \\
2005 Dec 30 & 53734.65 & 0.02 & 80 & 76.04	&$\zeta$ Pup, $\chi$ Car			& 10 \\
2005 Dec 31 & 53735.66 & 0.03 & 80 & 75.97	&$\zeta$ Pup, $\chi$ Car			& 16 \\

2006 Jan 01 & 53736.62 & 0.05 & 80 & 76.22	&$\zeta$ Pup, $\chi$ Car			&  9 \\
2006 Jan 02 & 53737.62 & 0.06 & 80 & 76.04	&$\zeta$ Pup, $\chi$ Car, J Pup		& 15 \\
2006 Jan 07 & 53742.64 & 0.12 & 80 & 76.30	&$\zeta$ Pup, $\chi$ Car			&  9 \\
2006 Jan 08 & 53743.62 & 0.14 & 80 & 76.28	&$\zeta$ Pup, $\chi$ Car			&  9 \\
2006 Jan 09 & 53744.65 & 0.15 & 80 & 76.08	&$\zeta$ Pup, $\chi$ Car, J Pup		& 13 \\

2006 Feb 08 & 53774.61 & 0.53 & 80 & 75.39	&$\zeta$ Pup, $\chi$ Car			& 10 \\
2006 Feb 10 & 53776.60 & 0.56 & 80 & 75.35	&$\zeta$ Pup, $\chi$ Car			& 15 \\
2006 Feb 11 & 53777.57 & 0.57 & 80 & 75.11	&$\zeta$ Pup, $\chi$ Car			& 24 \\
2006 Feb 13 & 53779.56 & 0.59 & 80 & 75.13	&$\zeta$ Pup, $\chi$ Car			& 21 \\

2006 May 11 & 53866.37 & 0.70 &  5 &  4.71	&$\zeta$ Pup, $\chi$ Car			&  5 \\
2006 May 11 & 53866.42 & 0.70 & 30 & 27.47	&$\zeta$ Pup, $\chi$ Car			&  4 \\

2006 Jun 17 & 53903.38 & 0.17 &  5 &  4.30	&$\zeta$ Pup, $\chi$ Car			&  7 \\
2006 Jun 18 & 53904.37 & 0.18 & 15 & 12.95	&$\zeta$ Pup, $\chi$ Car			&  6 \\

     \hline
    \end{tabular}
 \end{minipage}    

\end{table*}

%
%

\section{Orbital Solution}
\label{fit}

The theoretical response of a two aperture interferometer to the combined 
light of a binary star can be given by \citep{HBrown70}
\begin{equation}
\label{binary_v2}
V^2 = \frac{V_1^2 + \beta^2V_2^2 + 2\beta |V_1||V_2| \cos(2\pi\bmath{b} \cdot \brho/\lambda)}
      { ( 1 + \beta ) ^2 },
\end{equation}
where $V_1$, $V_2$ are the visibilities of the primary and secondary 
respectively and $\beta < 1$ is the brightness ratio of the two stars in the 
observed bandwidth. The angular separation vector of the secondary with 
respect to the primary is given by $\brho$ (measured East from North),
$\bmath{b}$ is the baseline vector projected onto the plane of the sky and 
$\lambda$ is the centre observing wavelength. The observed $V^2$ will vary 
throughout the night due to Earth rotation of $\bmath{b}$ and the orbital 
motion of the binary. The Keplerian orbit of a binary star, i.e. $\brho$ as 
a function of time, can be parameterized with seven elements: the period 
$P$, semi-major axis $a$, eccentricity $e$, epoch of periastron $T_0$, the 
longitude of periastron $\omega$, the longitude of ascending node $\Omega$ 
and the inclination $i$. When using two-aperture optical interferometry, the 
phase of the complex visibility is lost and hence, $\omega$ and $\Omega$ 
have an ambiguity of 180\degr. This stems from the fact that the identity of the 
components cannot be determined. Radial velocity measurements can be used to remove the ambiguity of 
$\omega$ but that of $\Omega$ remains. Interferometers with three or more 
apertures can use closure phase techniques to identify the components and 
hence remove the 180\degr ambiguity of $\Omega$. 

\subsection{Component Visibilities}

In the simplest case, stars can be modeled by a disc of uniform irradiance 
with angular diameter $\theta$. The visibility is then given by
\begin{equation}
\label{udisc_v}
V = \frac{2 J_1(\pi |\bmath{b}| \theta / \lambda)}
            {\pi |\bmath{b}| \theta / \lambda},
\end{equation}
where $J_1$ is a first order Bessel function. Real stars are 
limb-darkened, therefore corrections to the uniform disc diameter are needed 
to find an estimate of the true angular diameter. In the case of a compact 
atmosphere, the corrections are small but this may not be the case with 
extreme-limb darkened stars or stars with extended atmospheres 
\citep{Tango02}. In the case of emission line stars, care must be taken to 
ensure that only the continuum is observed (i.e. free of line emission) 
because a layer that forms an emission line cannot be assumed to be at the 
same radius as the continuum. The optically thick stellar wind of WR stars 
not only produces emission lines but also obscures the hydrostatic core 
surface \citep{Moffat96}. 

Over SUSI's wide observing band, emission lines contributed a significant 
portion of the detected light. The expected spectral response of SUSI 
(the combination of the 700\,nm filter profile and fig. 5 of
\citealt{DeMarco00}) was used to 
estimate that four emission lines contribute approximately 32 percent of 
the light received from the WR component during our observations. Three 
lines are blends of HeI--CIII, CIII--HeI and CII--CIII at 
$\lambda\lambda$674.1, 706.6, 723.0\,nm respectively. The remaining line is 
due to HeII at $\lambda$656.0\,nm. 

As each emission line corresponds to a different excitation/ionisation,
it follows that each line could be formed at a different radius in the 
stellar wind. Moreover, emission lines may form over a range of radii.
\citet{Hillier89} and \citet{Dessart00} have used WR models to predict the 
line formation stratification in the stellar wind. \citet{Niedzielski94} and
\citet*{SLadbeck95} have shown that observed WR emission lines have differing 
equivalent widths and concluded that these differences were evidence of 
excitation/ionisation stratification within the WR stellar wind. 
In the case of $\gamma^2$~Vel, \citet{HBrown70} estimated the uniform disc 
angular diameter of the $\lambda465.0$\,nm CIII emission region to be 
$2.05\pm0.19$\,mas and the analysis of the VLTI included modeling of the 
relevant WR IR emission line layers as uniform discs, each with a small 
difference in diameter \citep{Millour06}.

Unfortunately, the radii of the detected emission line forming layers 
have not been measured over SUSI's optical range and SUSI's red system 
presently has no way of isolating a single emission line. One possibility 
is that the four detected emission line forming layers are at essentially 
equal radii allowing a simple single uniform disc model of the emission. 
However, a Gaussian irradiance profile may be more applicable if the layers are
spread in radii.

We therefore analyse the interferometric data using two models of the WR 
star, each with a continuum core as a disc of uniform irradiance with 
angular diameter $\theta_{\rm C}$. The emission line layers are modeled to 
be either a single uniform disc of angular diameter $\theta_{\rm U}$ or 
a Gaussian shell with angular full-width half-maximum of 
$\theta_{\rm G}$ contributing $I_{\rm E}$ of the detected irradiance. The 
theoretical visibilities for these models are:
\begin{equation}
\label{wr_m1}
V_{\rm WR,U} = (1-I_{\rm E})\frac{2 J_1(\pi |\bmath{b}| \theta_{\rm C} / \lambda)}
                         {\pi |\bmath{b}| \theta_{\rm C} / \lambda} +
           I_{\rm E} \frac{2 J_1(\pi |\bmath{b}| \theta_{\rm U} / \lambda)}
                         {\pi |\bmath{b}| \theta_{\rm U} / \lambda}
\end{equation}
\begin{equation}
\label{wr_m2}    		
V_{\rm WR,G} = (1-I_{\rm E})\frac{2 J_1(\pi |\bmath{b}| \theta_{\rm C} / \lambda)}
                         {\pi |\bmath{b}| \theta_{\rm C} / \lambda} +
         I_{\rm E} \exp \left[
		    \frac{-\pi^2}{4\ln2}\frac{\theta_{\rm G}^2 |\bmath{b}|^2}{\lambda^2}
                 \right]
\end{equation}
The O-star primary is modeled as a uniform disc of angular 
diameter $\theta_{\rm O}$ as per equation (\ref{udisc_v}). Note that the 
brightness ratio $\beta$ in equation (\ref{binary_v2}) is usually given
in reference to the continuum irradiance detected in the observing band 
i.e.\ free from emission lines. As our observing band is contaminated with 
emission lines it is convenient to introduce the relative brightness 
$\beta'$ of the WR component (including emission lines) to that of the 
O-star in the observing band. The simple relation
\begin{equation}
\label{beta_p}
\beta = (1-I_{\rm E})\beta'
\end{equation}
can then be used to estimate the usual continuum brightness ratio.


\subsection{Fitting Procedure and Uncertainty Estimation}
\label{fit_err}

Equation (\ref{binary_v2}) is strictly only valid for observations over
very narrow bandwidths. For real detection systems
{\em wide bandwidth effects} can reduce the observed $V^2$. 
\citet{Tango02} present corrections which can be applied to wide bandwidth
interferometric observations of single stars. In the case of binary stars,
the way in which the wide bandwidth affects the observed squared visibility
($V^2$) is dependent on the detection system and subsequent calculations
\citep{Boden99}. For a scanning detection system (such as the one used
at SUSI) the modulating term in equation (\ref{binary_v2}) will be
reduced by a factor that is dependent on the separation of the component 
phase centres in delay space and the spectral response of the 
interferometer. Approximating the spectral response as a Gaussian of centre 
wavelength $\lambda_0$ with full-width half-maximum $\Delta\lambda$ and for 
convenience defining $\psi = 2\pi\bmath{b} \cdot \brho/\lambda_0$, then the 
equivalent to equation (\ref{binary_v2}) giving $V^2$ for a binary star
for the case of a wide spectral bandwidth is:
\begin{equation}
\label{wide_v2}
V^2 = \frac{V_1^2 + \beta^2 V_2^2 + 
	    2\beta r(\psi)|V_1||V_2| \cos(\psi)}
    	    { ( 1 + \beta ) ^2 },
\end{equation}
where
\begin{equation} 
r(\psi) = \exp \left[\frac{-\Delta\lambda^2}{\lambda_0^2}\frac{\psi^2}{32\ln2}\right].
\end{equation}
The term $r(\psi)$ corresponds to the autocorrelation of the Gaussian
envelope of the interference pattern. As $\Delta\lambda$ approaches zero, 
i.e. a narrow bandwidth, then equation (\ref{wide_v2}) reduces to equation 
(\ref{binary_v2}). 

In the case of $\gamma^2$~Vel, the emission lines (in an otherwise 
near-Gaussian spectral response determined by the 700\,nm filter profile
and fig. 5 of \citealt{DeMarco00}) alter the shape of the modulated 
interference fringes, further complicating the estimation of $V^2$. Detailed 
numerical simulations showed that the error introduced when approximating 
the spectral response as Gaussian was smaller than the associated 
measurement error. Hence a Gaussian approximation of the spectral response 
was deemed adequate.
 
Initial values of $i$ and $\Omega$ were found via a coarse search of parameter 
space with the remaining orbital parameters limited to within 3 standard
deviations of the values given in \citet{Schmutz97} and the stellar model 
parameters ($\theta_{\rm O}$, $\theta_{\rm C}$, $\theta_{\rm G}$/$\theta_{\rm U}$, $\beta$, $I_{\rm E}$)
were estimated using \citet{HBrown70}, \citet{Schmutz97} and 
\citet{DeMarco00}.

The final estimation of parameters was completed using $\chi^2$ minimization
as implemented by the Levenberg-Marquardt method to fit equation 
(\ref{wide_v2}) to the observed $V^2$. As the radial velocity measures have 
occurred over a long time frame of many periods, the spectroscopically 
determined period will be of higher accuracy than that determined by 
interferometry. Therefore the period given in \citet{Schmutz97} was adopted 
and the remaining orbital parameters were allowed to vary. The angular 
diameters were initially fixed to 0.44\,mas for the primary 
(from \citealt{HBrown70}) and 0.22\,mas for the WR continuum core 
i.e.\ a continuum\footnote{The continuum radius corresponding to a mean 
Rosseland optical depth of approximately unity is given in \citet{Schmutz97}
to be about 6\,R$_{\sun}$.} radius of 6\,R$_{\sun}$  at the {\em Hipparcos} 
distance of 258\,pc. The emission line contribution for each model was fixed 
to 32 percent of the received WR star irradiance and the brightness ratio 
$\beta'$ was a free parameter. These values are not critical as the resolution 
of the SUSI configuration during observations is such that the WR continuum 
is essentially unresolved. Moreover $V_1$, $V_2$ \& $\beta'$ (and hence
$\theta_{\rm O}$, $\theta_{\rm C}$, $\theta_{\rm G}$/$\theta_{\rm U}$ 
\& $I_{\rm E}$) are coupled and only affect the estimated orbital solution 
from equation (\ref{wide_v2}) if highly inappropriate values are adopted. 

When finding the minimum of the $\chi^2$ manifold, the non-linear fitting
program calculates the inverse of the covariance matrix. The diagonal 
elements are used to derive the formal uncertainties of the fitted parameters.
As equation (\ref{wide_v2}) is non-linear and the visibility measurement
errors may not strictly conform to a normal distribution, the formal 
uncertainties may be underestimates. To confirm the accuracy of the values 
derived from the covariance matrix, three uncertainty estimation methods 
were adopted.

Firstly, the model visibilities of each observation were subjected to Monte
Carlo realisation of the measurement error distribution to produce synthetic
data sets. Secondly, the {\em bootstrap} method \citep{Press92} was 
employed to produce a second population of synthetic data. By randomly
sampling {\em with replacement}, a synthetic data set is formed. The bootstrap
method has the advantage that the measurement error distribution is not
assumed to be known as per Monte Carlo but may not produce a representative
sample. The synthetic data sets from each method were used to estimate
the model parameters with the same non-linear fitting program, thus building a
distribution of each model parameter. 

The final uncertainty estimation method involves a likelihood based
random walk through parameter space using a Markov chain Monte Carlo (MCMC) 
simulation implemented with a Metropolis-Hastings algorithm 
(see Chapter 12 of \citealt{Gregory05a} for an introduction to MCMC). 
This approach yields the full marginal posterior probability
density function (PDF) but may fail to fully explore pathologically narrow
probability peaks in a reasonable number of iterations \citep{Gregory05b}. 
Furthermore, the current knowledge of the system can be included in the
analysis by assuming an {\em a priori} distribution. For example, the effect 
of adopting the spectroscopic period (and associated uncertainty) on the
remaining orbital parameters can be included into the uncertainty estimation
of the remaining model parameters. 

\subsection{Results}
\label{res}

\begin{table}
    \caption{Comparison of the fitted parameters of this work with those found
	     in the literature.}
    \label{fit_table}
    \begin{tabular}{@{}ccccc@{}}
	\hline
Parameter 	& Unit &   This Work     	& Literature       & Ref \\
	\hline
$P$ 	    	& days 	& $(78.53\pm0.01)^a$ 	& $78.53\pm0.01$   & S97  \\
$a\arcsec$    	& mas  	&   $3.57\pm0.05$    	&  $4.3\pm0.5$     & HB70 \\
$e$ 	        &      	& $0.334 \pm0.003$   	&  $0.326\pm0.01 $ & S97  \\
$T_0$       	& MJD  	& $50120.4\pm0.4$    	&  $50120\pm2$     & S97  \\
$\omega$    	& deg  	&   $67.4 \pm 0.5$   	& $68\pm 2 $       & S97  \\
$\Omega$    	& deg  	&  $247.7 \pm 0.4$   	& $[-80, -20]^{bc}$& SL87 \\
		&      	&                    	& $142^b$          & S97  \\
$i$         	& deg  	&    $65.5 \pm 0.4$  	& $65\pm 8 $       & S97  \\
	        &      	&		    	& $63\pm 3 $       & DM00 \\
$\theta_{\rm O}$& mas	& $0.47\pm0.05$      	& $0.44\pm0.05$    & HB70 \\
$\theta_{\rm C}$& mas	& $(0.17\pm0.08)^a$  	&      -	   &  -   \\
$\theta_{\rm G}$& mas	&  $1.3 \pm 0.3$     	&      -	   &  -   \\
$\theta_{\rm U}$& mas	&   $1.7\pm0.2$      	&      -	   &  -   \\
$\beta'$        &     	&  $0.45 \pm0.02$    	&      -           &  -   \\
$I_{\rm E}$     &      	& $(0.32 \pm 0.03)^a$	&    $0.32$        & DM00 \\
	\hline
    \end{tabular}
    \newline
    $^a$ adopted parameters, $^b$ definition is ambiguous, see text for 
    more details, $^c$ square brackets denote a range.
    \newline
    HB70: \citet{HBrown70}; SL87: \citet{StLouis87}; S97: \citet{Schmutz97};
    DM00: \citet{DeMarco00}
\end{table}

Initial analysis resulted in orbital parameters that, when combined with
the spectroscopic values as shown in Section~\ref{dist}, produced a distance 
of approximately 330\,pc. Consequently, the angular diameter of the WR 
continuum core was revised to 0.17\,mas (consistent with radius expectations
at this new distance) which had a negligible small effect the fitted 
orbital parameters. We note that the WR core is essentially unresolved
and therefore any diameter uncertainty (such as limb-darkening corrections 
to the equivalent uniform disc angular diameter) are insignificant.

\begin{figure}
 \includegraphics[width=\linewidth]{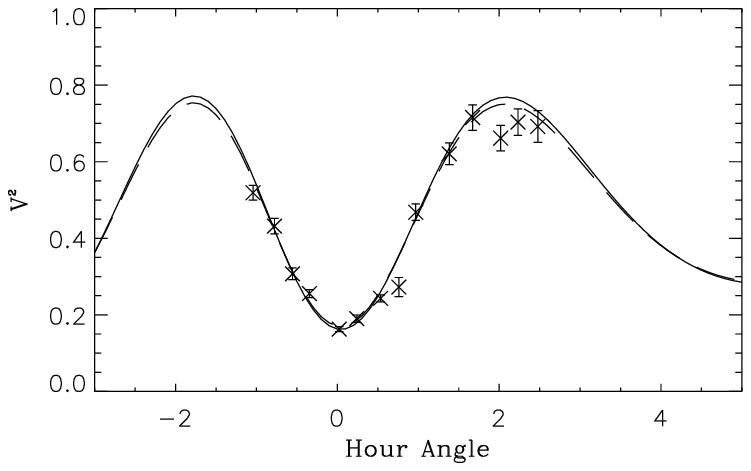}
 \includegraphics[width=\linewidth]{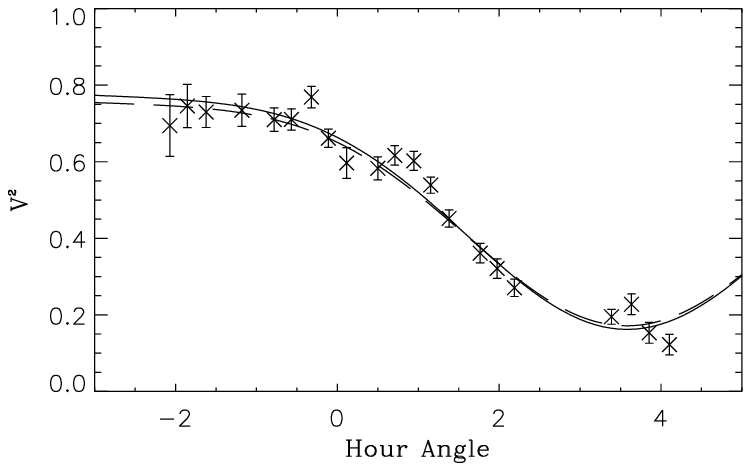}
 \caption{Data from the nights of 2005 Mar 11 and 2006 Feb 13 where
 each data point represents a measure of $V^2$ with the associated formal
 error. The values of Table~\ref{fit_table} have been been used
 to show the uniform disc (solid) and Gaussian shell (dashed) 
 emission layer models.}
 \label{fig:night_example} 
\end{figure}

\begin{figure}
 \includegraphics[width=\linewidth]{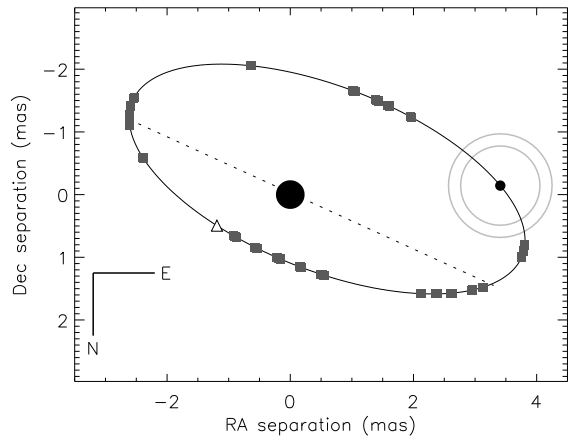}
 \caption{The relative orbit (solid line) of the WR-star about the O-star
 projected on the plane of the sky at an orbital phase of 0.4. The dotted 
 line is the line-of-nodes, the open triangle signifies periastron and the 
 grey squares are parts of the orbit that were observed with SUSI. The 
 stellar components (filled circles) are drawn to scale using the values of 
 Table~\ref{fit_table}. The full-width half-maximum of the Gaussian shell 
 and the angular diameter of the uniform disc emission line layer models are 
 shown as the grey circles about the WR component.}
 \label{fig:orbit} 
\end{figure}

The results of the two models (uniform disc and Gaussian shell) were 
completely consistent in all orbital parameters and the final values of the 
fitted parameters are given in Table~\ref{fit_table}. Example data from two 
nights with the fitted models is shown in Fig.~\ref{fig:night_example}
and the projected orbit on the plane of the sky is shown in 
Fig.~\ref{fig:orbit}. The 
reduced $\chi^2$ of the two fits were approximately 2.2 implying that the 
formal measurement errors calculated by the data reduction software are 
underestimated by 49 percent. Therefore the measurement uncertainties were 
scaled by 1.49 before the parameter uncertainties were estimated using 
the techniques outlined in Section~\ref{fit_err}. The Monte Carlo and 
bootstrap methods were set to each generate $10^3$ synthetic data sets 
while the MCMC simulations completed $10^7$ iterations. All combinations 
of model and uncertainty estimation methods produced similar distributions 
of the free parameters which were, for the orbital elements, Gaussian in 
appearance. Furthermore, the standard deviation of each model parameter 
from all uncertainty methods were consistent to within rounding. The 
standard deviations derived from the MCMC simulations were only slightly 
larger as the probability space of all parameters was explored: the period, 
WR continuum core diameter and emission line contribution varied within a 
Gaussian likelihood with assumed values given in Table~\ref{fit_table}. The 
distributions of the emission region parameters were somewhat asymmetric 
with a larger tail at higher values. 
This is a direct result of the small number of short baseline measurements 
combined with the coupling of the component visibilities and the brightness 
ratios. All uncertainties quoted in Table~\ref{fit_table} are the standard 
deviation values of the MCMC simulations as we believe they represent the 
most realistic and conservative parameter uncertainty estimates for our 
data set. 

There is no evidence that the visual companion $\gamma^1$ Vel
or the nearby K4~V star detected by \citet{Tokovinin99} affect our data.
$\gamma^1$ Vel is approximately 2.5 magnitudes fainter and separated by
41\arcsec\, from $\gamma^2$~Vel; the K4~V star has a separation from
$\gamma^2$~Vel of approximately 4.7\arcsec\, and is approximately 14.8 
magnitudes fainter \citep{Tokovinin99}. These stars are either out of the 
field-of-view or too faint to be detected by SUSI. 

The orbital elements derived from radial velocity measurements 
\citep{Schmutz97} are consistent with our data as is the inclination inferred
by \citet{Schmutz97} and \citet{DeMarco00}. The discrepant angular semi-major
axis of \citet{HBrown70} is largely due to their adoption of the inclination 
($a$ and $i$ are co-dependent as shown in \citealt{Hummel93}). This 
correlation between $a$ and $i$ was confirmed by the ad-hoc fixing of the 
inclination as per \citet{HBrown70} and the resultant angular semi-major 
axis was found to be consistent with their value. 

The single angular separation vector of \citet{Millour06} could be used to 
remove the 180\degr ambiguity in our value of $\Omega$. However, \citet{Millour06} 
have stated their belief that they have identified the components correctly 
but are cautiously awaiting the final analysis of calibration data. 
Therefore, we tentatively set the position angle of the ascending node to 
that given in Table~\ref{fit_table}. Using the final orbital values at the 
epoch of the VLTI measurement, the separation of the components is 
$3.85\pm0.06$ \,mas and the WR component is located $80\pm1\degr$ (measured 
Eastwards from North) relative to the O-star. These values are marginally 
consistent with the VLTI measurement of $3.62^{+0.11}_{-0.30}$\,mas and 
$73^{+9}_{-11}$\,deg given in \citet{Millour06}.

\subsection{Polarimetry Analysis}
\label{pol}

Polarimetric analyses of $\gamma^2$~Vel have produced the only pre-existing
estimations of the position angle of the ascending node. However, 
the parameter, $\Omega$, quoted in the polarimetry literature is {\em not}
the position angle of the ascending node but rather the position angle of 
the rotation axis (orbit normal) projected on to the plane of the sky 
\citep*{BME78}. This angle is included in the polarimetry 
analysis to account for the rotation of the binary system in relation to the 
coordinate system of the observed Stokes parameters $Q$ and $U$.
In the plane of the sky, the relationship between the two definitions of 
$\Omega$ is a simple difference of 90\degr. However, the polarimetry 
rotation parameter $\Omega$ is usually defined in the $Q$-$U$ plane (see
\citealt{Robert89}) and hence an additional factor of two is introduced i.e.\
\begin{equation}
    \label{eq:pol_Om}
    \Omega_{\it an} = \frac{\Omega_{\it ra}}{2} \pm 90
\end{equation}
where $\Omega_{\it an}$ and $\Omega_{\it ra}$ are the position angles of the
ascending node in the plane of the sky and the projected rotation axis  
in the $Q$-$U$ plane respectively. The addition or subtraction of 90\degr\,
is dependent on the orientation of the binary orbit with respect to the North 
Celestial pole. Unfortunately neither \citet{StLouis87} nor \citet{Schmutz97}
give details of which plane their quoted $\Omega_{\it ra}$
is measured in. For the range given in \citet{StLouis87} $\Omega_{\it an}$
could be 190\degr--250\degr or 230\degr--260\degr if transformed from the plane of the sky or from 
the $Q$-$U$ plane respectively. Thus, the values of \citet{StLouis87} are 
compatible with the interferometric value regardless of which plane their 
$\Omega$ value is measured. The $\Omega_{\it an}$ value of \citet{Schmutz97} 
could be 232\degr (plane of the sky) or 251\degr ($Q$-$U$ plane) respectively. 
Thus $\Omega_{\it an}$ derived from the $Q$-$U$ plane is close to our value. 
However, without an estimate of the uncertainty we cannot determine its 
consistency with our value.

As polarimetric data of $\gamma^2$~Vel are the only other available 
{\em experimental} values to estimate the position angle of the ascending
node and the inclination (and to update the polarimetric mass-loss rate 
determined by \citealt{StLouis88} in Section~\ref{massloss}) the polarimetry 
data of \citet{StLouis87} were reanalysed using the model of \citet{Brown82} 
(with the corrections made by \citealt{Simmons84}) and the rotation angle 
$\Omega_{\it ra}$ included in the $Q$-$U$ plane as per \citet{Robert89}. A 
$\chi^2$ minimization was performed with the period and eccentricity fixed 
to the interferometric values given in Table~\ref{fit_table}. The starting 
values for the remaining parameters were found using the interferometric 
values given in Table~\ref{fit_table}. The best fit values were 
$T_0 = 50120.9$ MJD, $\lambda_p = 163\degr$, $\Omega_{\it ra} = 303\degr$, 
$i=66\degr$, $\tau_* = 0.036$, $Q_0 = 0.077$ and $U_0 = -0.077$. Converting 
to usual binary star parameters, $\lambda_p = \omega + 90\degr$ and 
equation (\ref{eq:pol_Om}) give $\omega = 73\degr$ and 
$\Omega_{\it an} = 242\degr$. A $10^7$ iteration MCMC simulation was 
performed to investigate the effect that the assumed parameter and 
eccentricity values have on those that were fitted. All parameters with the 
exception of the inclination produced Gaussian PDFs centred roughly on the 
best fit values. However, the PDF of the inclination had multiple peaks with 
90 percent of the probability contained between 64.5\degr and 77.5\degr. 
This confirms the statement by \citet{Schmutz97} that the inclination cannot 
be accurately constrained by the available polarimetry data. The PDF of 
$\Omega_{\it an}$ produces an uncertainty of approximately 4\degr and hence 
our polarimetric value of $\Omega_{\it an}$ is consistent with the 
interferometrically determined value.

%
%

\section{Discussion}
\label{disc}

\subsection{Wind-Wind Interactions}
\label{windwind}

The orbital values of Table~\ref{fit_table} indicate that the angular 
separation at periastron is approximately 1.29\,mas suggesting that the wind
of the O-star primary may interact with the emission line forming layers 
of the WR wind. Moreover, a higher temperature and density colliding
winds zone (as inferred by the X-ray spectrum of $\gamma^2$~Vel 
\citealt*{Willis95}; \citealt{Stevens96}; \citealt*{Henley05})
may also affect the appearance of the emission line forming layers.
However, the angular diameter of these layers, whether 
modeled as a uniform disc or Gaussian irradiance distribution, is not 
adequately constrained by our data. The effect of an irradiance asymmetry 
due to wind-wind interactions on the orbital parameters is considered to 
be negligibly small as the WR wind is expected to dominate that of the 
O-star. Furthermore, the short baseline data, which should be most strongly
affected, are limited and were collected during an orbital phase when the 
O-star was outside the estimated emission line forming layers. 

\subsection{Distance}
\label{dist}

The distance to a double-lined spectroscopic binary star that has a 
well-determined visual or interferometric orbit can be found using the 
dynamical parallax \citep{Heintz78}:
\begin{equation}
\label{dyn_par}
\pi_d = \frac{a\arcsec}{(a_1 + a_2)}.
\end{equation}
The semi-major axis of the primary (secondary) component orbit
about the system centre-of-mass, $a_1$($a_2$), is measured
in astronomical units. 

Using the semi-amplitude
of the component radial velocity given in \citet{Schmutz97}, 
($K_1=38.4\pm2$\,kms$^{-1}$, $K_2=122\pm 2$\,kms$^{-1}$) and
the interferometric orbital values, $a_1$ and $a_2$ were calculated 
(in km) using
\begin{equation}
a_{1,2} = \frac{43200 K_{1,2} P \sqrt{1-e^2}}{\pi\sin i}.
\end{equation}
These values are given in Table~\ref{phys_table}. The resulting dynamical 
parallax is $\pi_d = 2.97\pm0.07$\,mas which yields a distance of 
$336^{+8}_{-7}$\,pc. 

Compared to the {\em Hipparcos} (\citealt{Schaerer97}; \citealt{Hucht97}) 
and the VLTI distance \citep{Millour06} our uncertainty in the measured
distance is lower by an about order of magnitude. 
Moreover, there is considerable contrast between the distance 
determinations. It has been shown that the dynamical parallax of some binary
stars can be significantly different to the {\em Hipparcos} parallax
(\citealt{Davis05}; \citealt{Tango06}). Binary motion can influence the
{\em Hipparcos} value and where possible, has been included in the
data reduction \citep{Lindegren97}. The $\gamma^2$~Vel entry in the 
{\em Hipparcos} catalogue \citep{ESA97} indicates that 
it was not treated as binary but it is entered into the {\em Variability
Annex: Unsolved variables}. Furthermore, a critical reassessment
of the {\em Hipparcos} catalogue has identified a number of defects
in the data \citep{Leeuwen05a} and a new reduction of the raw data is
in progress \citep{Leeuwen05b}. The preliminary revision to the {\em Hipparcos} 
parallax of $\gamma^2$~Vel is $3.35 \pm 0.29$\,mas (van Leeuwen, private communication) 
i.e.\ a distance of $299^{+28}_{-24}$\,pc which is marginally consistent
with our value. The VLTI distance \citep{Millour06} relies on 
a single angular separation measure and assumed orbital elements 
from the literature which are inferior to our interferometric values. 

The $\gamma^2$~Vel {\em Hipparcos} parallax raised the question of the 
assumed membership of the Vela OB2 association due to the disparity of their
distances: 258\,pc vs 450\,pc. The {\em Hipparcos} proper motion and 
parallaxes of nearby OB associations including Vela OB2 were analysed by 
\citet{deZeeuw99} who concluded that $\gamma^2$~Vel was an edge member with 
probability of 64 percent. The 93 members produce a mean distance of 
$410 \pm 12$\,pc \citep{deZeeuw99} i.e.\ a mean parallax of approximately 
2.44\,mas. Using fig. 11 of \citet{deZeeuw99} it was estimated that 60 
percent of the Vela OB2 members lie within 0.5\,mas of the mean value. 
Furthermore, \citet{Pozzo00} model the {\em Hipparcos} parallax dispersion 
as a Gaussian of standard deviation 0.68\,mas. Our parallax moves 
$\gamma^2$~Vel from the edge of Vela OB2 to well within the association. 
Using the Galactic coordinates of $\gamma^2$~Vel in \citet{Hucht01} (and a
Sun-Galactic centre distance of 8\,kpc) the 
revised Galactocentric distances are $R_{\rm GC} = 7.97$\,kpc and 
$z_{\rm GC}=-45$\,pc. We estimate an uncertainty in $R_{\rm GC}$ of 
0.01\,kpc and $z_{\rm GC}$ of 1\,pc which are a factor of four less than 
those estimated with the Hipparcos distance. Compared to values found in 
the literature ($R_{GC} = 8$ kpc, $z_{GC}=-35$\,pc, \citealt{Hucht01}) our 
values place $\gamma^2$~Vel farther from Galactic plane but at approximately 
the same distance from the Galactic centre.

\subsection{Component Magnitudes}

The absolute visual magnitude of the system and component stars can be 
redetermined using our distance. There is the possibility of broadband
photometry (\citealt{Johnson66}; \citealt{Cousins72}) suffering 
contamination by the WR emission lines (\citealt{Smith68b};
\citealt{DeMarco00}) and studies of galactic WR distributions and 
intrinsic parameters almost exclusively use narrowband `line-free'
photometry (e.g.\ \citealt{Smith68b}; \citealt{Conti83}; \citealt{Conti90}).
The mean and standard deviation of narrowband apparent magnitude and 
extinction values found in the literature are $v=1.73 \pm 0.03$\,mag 
(\citealt{Hucht88}; \citealt*{Smith90}; \citealt{Conti90};
\citealt{Hucht97}; \citealt{DeMarco00}) and $A_v =0.09 \pm 0.06$\,mag 
(\citealt{Conti90}; \citealt{DeMarco00}; \citealt{Hucht01}).
Therefore, combining these values with our distance we obtain the system 
absolute magnitude of $M_v({\rm WR+O}) = -5.99$. 

The component absolute magnitudes in both narrowband ($v$) and broadband 
($V$) can then be found using the component brightness ratio in $v$ and 
$M_v - M_V = -0.1$\,mag \citep{Hucht01}. The output brightness ratio $\beta'$
from the fitting procedure and the estimated emission line contribution 
$I_{\rm E}$ produce a continuum brightness ratio in our observing band of 
$\beta = 0.31 \pm 0.03$. This is consistent with the expected brightness 
ratio at 700nm of $0.35\pm0.04$ determined from fig. 1 of 
\citet{DeMarco00}. We therefore adopt the $v$ brightness ratio of 
\citet{DeMarco00} rather than scaling the continuum brightness ratio in our 
observing band. We obtain $M_v$(WR)$=-4.33\pm0.17$\,mag and
$M_V$(O)$ = -5.63\pm0.10$\,mag (from $M_v$(O)$ = -5.73\pm0.10$\,mag) where 
the estimates of the magnitude uncertainties include the distance, reddening 
and the component brightness ratio. The uncertainties in previous estimates of 
the distance to $\gamma^2$~Vel have dominated error determinations of 
the component absolute magnitudes (\citealt{Hucht97}). 
This limitation has now been removed with our measurement.

\subsection{O-star Component}
\label{Ostar}

The most recent classification of the O-star spectral type is O7.5 given 
by \citet{DeMarco99}. They also deduce a bright-giant luminosity class (II) 
from analysis of their {\em synthetic} HeII line $\lambda468.6$\,nm as other 
lines are unavailable due to assumed metallicity in their model and blending 
effects prevent an observational determination. The recent spread-spectrum 
interferometric analysis of \citet{Millour06} has found the O-star models 
that best fit their data have gravity values more in line with a super-giant 
classification. However, the {\em Hipparcos} distance produces 
$M_V$(O)$ \simeq -5.1$\,mag which is more consistent with a giant or dwarf 
hence the O7.5~III--V classification in the {\em VIIth catalogue of galactic
Wolf-Rayet stars} \citep{Hucht01}. Moreover, the calibration of physical 
parameters of O-stars has evolved as stellar models develop and clear 
separation into only three luminosity classes I, III \& V has become 
standard. Assuming a spectral type of O7.5, our $M_V$(O) is consistent with 
a giant classification using \citet{Howarth89} or \citet*{Vacca96} but falls 
between a giant and a super-giant with the calibration of \citet*{Martins05}.
Therefore a luminosity class of II--III for the O-star seems appropriate 
until the knowledge of this area of the HR diagram improves. 

The radius of the O-star is obtained from the measured angular diameter
and distance to the system was found to be $17\pm2$\,R$_{\sun}$. This is in 
excellent agreement with the earlier measurement by \citet{HBrown70}. A 
radius of $17$\,R$_{\sun}$ falls between all giant and super-giant
calibrations found in the literature (\citealt{Howarth89}; 
\citealt{Vacca96}; \citealt{Martins05}) supporting a reclassification of the
O-star to O7.5~II--III.

It is more appropriate to use an upper and lower estimate of the bolometric 
correction to find the O-star luminosity. Using \citet{Martins05} to place 
an upper (lower) limit to the bolometric correction (BC) of -3.4(-3.1), we 
obtain $\log L$(O)$/L_{\sun} = 5.51 (5.39)$. Using a mean luminosity of 
$\log L$(O)$/L_{\sun} = 5.45$ the age of the O-star is estimated at 
$(3.5 \pm 0.4) \times 10^6$\,yr from the single-star evolutionary models of 
\citet{Meynet94} (assuming Z=0.02) and an effective temperature of 
$T_{\it eff} = 35\,000$K \citep{DeMarco00}. However, rejuvenating
mass transfer (with WR-star as donor) may have occurred in an earlier
interactive phase. Although in good agreement with the estimates of 
\citet{Schaerer97} and \citet{DeMarco99}, this age should be considered a 
lower estimate. 
The age that we have estimated for the O-star and our re-confirmed 
membership of the Vela OB2 association implies that the
nearby population of low-mass, pre-main sequence stars detected by
\citet{Pozzo00} is coeval with $\gamma^2$~Vel. This population has
an age range estimated to be 2--6\,Myr \citep{Pozzo00}.

\subsection{WR-star Component}
\label{WRstar}

The expected $M_v$ of a WC8 star has previously been given as $-4.8$ to 
$-6.2$\,mag (\citealt{Smith68b}; \citealt{Conti83}; \citealt{Hucht88}) but 
more recently \citet{Hucht01} has estimated $-3.74$\,mag with a standard
deviation of 0.5\,mag. However, 
only three Galactic WC8 stars have a distance of sufficient certainty to be 
used in the current estimation of $M_v$ and previous studies relied almost 
exclusively on $\gamma^2$~Vel. We therefore individually compare our value
to the two other Galactic WC8 stars with distance determinations. 
WR135 has $M_v= -4.24$ and WR113 has $M_v= -3.68$
but is a binary star with persistent dust formation \citep{Hucht01}. Our 
value is close to that of WR135 and, as amorphous carbon dust emission is not
seen in the $\gamma^2$~Vel spectrum \citep{Hucht96}, the difference from that
of WR113 is not significant. 

Bolometric correction for a WC8 star is given in \citet*{Smith94} as -4.5
derived from evolutionary models and cluster membership. Recent
model BCs for $\gamma^2$~Vel are given in \citet{DeMarco00} and range
from -3.5 to -4.1. The atmospheric model analysis of the galactic WC8 single 
star WR135 by \citet{Dessart00} produces a BC = -4.0. Due to the non-binary 
nature of WR135 and its similarity to the WC8 component of $\gamma^2$~Vel,
we adopt a BC of -4.0. The resultant luminosity is 
$\log L$(WR)$/L_{\sun} \simeq 5.23$. This value is in complete accord with 
the luminosity determined by \citet{DeMarco00}.

\begin{table*}
  \centering
  \begin{minipage}{125mm}
    \caption{Physical parameters of $\gamma^2$~Velorum.}
    \label{phys_table}
    \begin{tabular}{@{}lcccc@{}}
	\hline
Parameter     & Unit	& This Work & Literature & Reference \\
	\hline
$a_1$         &$10^6$\,km & $43.0  \pm 2.2$   &	-	    &  -	 \\	
$a_2$         &$10^6$\,km & $136.6 \pm 2.3$   &	-	    &  -	 \\
$a_1$         & AU   	& $0.287 \pm 0.015$ &	-	    &  -	 \\	
$a_2$         & AU   	& $0.913 \pm 0.015$ &	-	    &  -	 \\
$\pi_d$       & mas  	& $2.97\pm0.07$     & $3.88\pm0.53$    & SSG97,H97  \\	
distance      & pc   	& $336^{+8}_{-7}$   & $350^{+50}_{-50}$, $258^{+41}_{-31}$, $368^{+38}_{-13}$ & HB70, SSG97/H97,M06  \\
        \hline
M$_{V}$(O)    & mag        & $-5.63\pm0.10$   &  $-5.10\pm0.10^a$         & DM00 \\
R(O)          & R$_{\sun}$ & $17\pm2$  	      & $17\pm3$, $12.4\pm1.7$  & HB70, DM99 \\        
L(O)          & L$_{\sun}$ & $2.8\times10^5$  & $(2.1\pm0.3)\times10^5$ & DM00 \\
${\cal M}$(O) & M$_{\sun}$ & $28.5\pm1.1$     & $30\pm2$                & DM00 \\        		
age(O)        & Myr 	   & $3.5\pm0.4$      &  $3.59\pm0.16$          & DM99 \\
	\hline	
M$_{v}$(WR) & mag    & $-4.33\pm0.17$ 	   &  $-3.84$                & DM00 \\
M$_{V}$(WR) & mag    & $-4.23\pm0.17$ 	   &  $-3.76\pm0.20^a$       & DM00 \\
R(WR)  & R$_{\sun}$  & $(6\pm3)^a$	   & $6$                     & S97 \\
L(WR)  & L$_{\sun}$  & $1.7\times10^5$     & $(1.7\pm0.4)\times10^5$ & DM00 \\
${\cal M}$ (WR)      & M$_{\sun}$          & $9.0\pm0.6$	& $9\pm2$   & DM00 \\
$\dot{\cal M}_{rad}$ & M$_{\sun}$\,yr$^{-1}$ & $3\times 10^{-5}$  & $3\times 10^{-5}$    &  SSG97\\
$\dot{\cal M}_{pol}$ & M$_{\sun}$\,yr$^{-1}$ & $8\times 10^{-6}$  & $7\times 10^{-6}$    &  S97\\
	\hline
    \end{tabular}
    \newline
    $^a$ neglected distance uncertainty.
    \newline
    SSG97: \citet{Schaerer97}; H97: \citet{Hucht97}; DM00: \citet{DeMarco99};
    DM00: \citet{DeMarco00};
    M06: \citet{Millour06}; HB70: \citet{HBrown70}; S97: \citet{Schmutz97}
    \end{minipage}
\end{table*}

\subsection{Stellar Masses}
\label{mass}

The component masses of a binary star can be extracted from the orbital
elements using Kepler's third law and the ratio of the component 
semi-major axes about the centre-of-gravity:
\begin{equation}
\label{kepler_mass}
{\cal M}_1 + {\cal M}_2  = \frac{(a_1 + a_2)^3}{P^2},
\end{equation}
\begin{equation}
\label{mass_ratio}
\frac{{\cal M}_1}{{\cal M}_2}  = \frac{a_2}{a_1}.
\end{equation}
When the semi-major axes are given in astronomical units and the period in 
years the resultant masses are in solar units. 
Using the values determined in Sections~\ref{res} and \ref{dist},
the masses of the $\gamma^2$~Vel components are 
${\cal M}$(O) = $28.5\pm1.1$\,M$_{\sun}$ and
${\cal M}$(WR) = $9.0\pm0.6$\,M$_{\sun}$. There is good agreement between 
our values and those derived from recent spectral analysis 
(\citealt{DeMarco99}; \citealt{DeMarco00}).

The mass of the O-star is consistent with a giant using the `spectroscopic'
calibration of \citet{Vacca96} but is a little low compared to the 
values of \citet{Howarth89}. A possible explanation for the higher value 
of \citet{Howarth89} is given in \citet{Vacca96} who note that evolutionary 
models produce masses systematically higher than spectroscopic models 
(the evolutionary model analysis in Section~\ref{Ostar} also produced a 
higher mass: $34\pm2$\,M$_{\sun}$). With the tables of \citet{Martins05} the 
O-star mass is consistent with the `theoretical' giant mass but is 
slightly high for the `observed' giant mass. Therefore our proposed 
luminosity classification of II--III is also supported by our observed mass 
and O-star calibrations in the literature.
 
The mass of the WR star determined with the mass-luminosity relationship 
given by equation 3 of \citet{Schaerer92} is 9.2\,M$_{\sun}$ (using the 
estimated $\log L$(WR)$/L_{\sun}$ from Section~\ref{WRstar}) which is in 
good agreement with our mass value. In the catalogue of \citet{Hucht01} 
there are 6 WC stars that have reliable masses and range from 
9--16\,M$_{\sun}$ with a mean of $12\pm3$\,M$_{\sun}$, consistent with our 
value. 

The mass and $M_v$(WR) we have determined can be used to check our choice
of BC in Section~\ref{WRstar}. Using fig. 4 of \citet{Smith94}, our mass and 
$M_v$ produces a point that lies above the -4.5 BC line but is lower than
the WC7 and WC9 points. This implies a bolometric correction between 
-3.5 and -4.5, justifying our choice of -4.0 from the atmospheric analysis
of WR135. 

\subsection{Mass-loss}
\label{massloss}
 
Previous observational mass-loss estimates of the WR star require adjustment 
using our distances and revised parameters before turning to a comparison 
with the expected values from theory. The mass-loss derived from 
radio continuum at 4.8\,GHz of \citet*{Leitherer97} with the wind velocity 
$v_{\infty}=1450$\,kms$^{-1}$ from \citet{Eenens94} and corrected for 
nonthermal emission by \citet{Chapman99} is 
$\dot{\cal M}_{\rm rad} \simeq (3.0 \pm 0.5) \times 10^{-5}$\,M$_{\sun}$\,yr$^{-1}$.
This is in agreement with the average radio-derived mass-loss
rate of WC8--9 stars of 
$\dot{\cal M}_{\rm rad} = (2 \pm 1) \times 10^{-5}$\,M$_{\sun}$\,yr$^{-1}$
determined by \citet*{Cappa04}. The polarimetric mass-loss of 
\citet{StLouis88} has been given more attention due
to the improved orbital parameters from the interferometric orbit. 
As described in Section~\ref{pol}, we reanalysed the polarimetric data of
\citet{StLouis87} but restricted all orbital parameters to the values given
in Table~\ref{fit_table} to obtain $\tau_* = 0.035$, $Q_0 = 0.077$ and 
$U_0 = -0.078$. This corresponds to a polarisation amplitude near periastron 
of $A_p \simeq 0.0009$. Once again using $v_{\infty}=1450$\,kms$^{-1}$ and 
\citet{StLouis88} the polarimetric mass-loss is found to be
$\dot{\cal M}_{\rm pol} \simeq (8\pm3)\times 10^{-6}$\,M$_{\sun}$\,yr$^{-1}$. 
The disparity of the radio and polarimetric measurements can be attributed 
to density variations
or clumping in the WR wind \citep{Moffat94}. Clumping affects the radio 
mass-loss as it assumes a smooth wind and is proportional to the wind 
density squared. On the other hand, polarimetric mass-loss is proportional 
to the density and is therefore insensitive to clumping 
(\citealt*{Nugis98}; \citealt{Hamann98}). The ratio of the 
above radio and polarimetric mass-loss gives an estimate of a clumping 
factor $m \simeq 3.8\pm1.8$. \citet{Moffat94} predict a value of $\ga 3$ and
the value for the single WC analysed by \citet{Hamann98} is 
$\sqrt{D} = m = 4$. \citet{Schmutz97} re-estimated 
$\dot{\cal M}_{\rm pol} \simeq 7 \times 10^{-6}$\,M$_{\sun}$\,yr$^{-1}$
and used the {\em Hipparcos} $\dot{\cal M}_{\rm radio}$ of 
\citet{Schaerer97} to find $m$ in the order of 4. The consistency
with our value is a fortuitous accident
of the polarimetric mass-loss changing by a small amount and the
radio mass-loss remaining unchanged at 
$\simeq 3 \times 10^{-5}$\,M$_{\sun}$\,yr$^{-1}$ -- the removal of the nonthermal 
component in the radio flux negated the revision of the distance.

Theoretical mass-loss rates for a 9\,M$_{\sun}$ (smooth stellar wind) WR
are $\simeq (1.3 \pm 0.4) \times 10^{-5}$\,M$_{\sun}$\,yr$^{-1}$ (\citealt{Doom88}; 
\citealt{Nugis00}). For the clumping-corrected, chemical composition
dependent relationship of \citet{Nugis00} we also find 
$\simeq (1.3 \pm 0.4)\times 10^{-5}$\,M$_{\sun}$\,yr$^{-1}$. 
These values are consistent within errors of both our mass-loss estimates. 
Further comparison can be made to the detailed modeling of
$\gamma^2$~Vel by \citet{DeMarco00} with the use of a clumping volume filling 
factor $f_{\rm v}$. Our clumping factor $m$ is related to $f_{\rm v}$ 
by $\sqrt{f_{\rm v}} = m^{-1}$ \citep{Hamann98} 
i.e. $f_{\rm v} \simeq 7 \%$ and gives 
$\dot{\cal M} \simeq 8 \times 10^{-6}$\,M$_{\sun}$\,yr$^{-1}$ in 
agreement with our polarimetric mass-loss rate.

The WR wind performance or transfer efficiency 
$\eta = \dot{\cal M}v_{\infty}c/L$ represents the ratio of radial momentum in 
the stellar wind to the radiation momentum and is $\la 1$ for winds that 
are within the single-scattering limit for a radiatively driven flow
\citep{Owocki99}. However, \citet{Lucy93} have shown $\eta \simeq 10$ is
possible for winds that are radiatively driven with multiple-scattering 
effects. A typical value of $\eta \simeq 10$ is given in 
\citet{Owocki99} while \citet{DeMarco00} approximate $\eta \simeq 7$ 
for $\gamma^2$~Vel assuming a 10\% clumping volume filling factor. 
The wind performance, using our polarimetric mass-loss and 
luminosity with $v_{\infty}=1450$ kms$^{-1}$ from \citet{Eenens94}, 
is $\eta \simeq 3.5$; consistent with values where multiple-scattering 
effects drive the WR wind and mass-loss \citep{Lucy93}.

%
%

\section{Conclusion}
\label{conc}

We present the first complete orbital solution for $\gamma^2$
Vel based on interferometric measurements with SUSI. Emission line 
contamination has been simply modeled and found not to affect the orbital 
solution.

In combination with the latest radial velocity measurements, the dynamical
parallax and distance to $\gamma^2$~Vel have been revised. The constraints we
have placed on the Galactic location of $\gamma^2$~Vel are the tightest to 
date by an order of magnitude and membership of $\gamma^2$~Vel in the Vela 
OB2 association has been confirmed. The contrast between our distance and 
previous measurements in the literature is such that all distance-dependent 
fundamental parameters require revision e.g. absolute magnitudes, mass-loss 
rates. Moreover, the formal uncertainty in the distance can now be included 
in the determination of fundamental parameters. 

Using calibrations found in the literature, the radius, absolute visual 
magnitude and mass of the component O-star are consistent with a luminosity 
classification of II--III. The age that we have estimated for the O-star 
falls at the centre of the 2--6 Myr range estimated by \citet{Pozzo00} for 
the nearby population of low-mass, pre-main sequence stars, implying coeval 
formation with $\gamma^2$~Vel.

We have shown that the narrow-band absolute visual magnitude and mass
of the WC8 component is consistent with other galactic WR stars of
similar type and theoretical mass-luminosity relationships.
The mass-loss rates determined by radio and polarimetric measurements have 
been revised to include new information in the literature and our orbital 
parameters. These values allowed an estimation of the clumping volume 
filling factor which is used to show agreement between recent atmospheric 
model analysis and our clumping insensitive mass-loss rate.

\section*{Acknowledgments}

This research has been jointly funded by The University of Sydney and the 
Australian Research Council as part of the Sydney University Stellar
Interferometer (SUSI) project. 
We wish to thank Brendon Brewer for his assistance with the theory and 
practicalities of Markov chain Monte Carlo Simulations. Andrew Jacob, 
Stephen Owens and Steve Longmore provided assistance during observations. 
Michael Scholz and Gordon Robertson supplied useful comments during 
discussion. The SUSI data reduction pipeline was developed by Michael 
Ireland. We also thank Floor van Leeuwen who kindly provided the
preliminary revision to the {\em Hipparcos} parallax and the referee
Douglas Gies for his constructive comments.
JRN acknowledges the support provided by a University of Sydney 
Postgraduate Award. This research has made use of the SIMBAD database,
operated at CDS, Strasbourg, France

\bsp

\label{lastpage}


\begin{thebibliography}{99}

\bibitem[\protect\citeauthoryear{Baschek \& Scholz}{1971}]{Baschek71} 
Baschek B., Scholz M., 1971, A\&A, 11, 83, 12, 322 

\bibitem[\protect\citeauthoryear{Boden}{1999}]{Boden99} 
Boden A.F., 1999, in Lawson P.R., ed, 
Principles of Long Baseline Stellar Interferometry

\bibitem[\protect\citeauthoryear{Brown, McLean \& Emslie}{Brown et al.}{1978}]{BME78}
Brown J.C., McLean, I.S. Emslie, A.G., 1978, A\&A, 68, 415

\bibitem[\protect\citeauthoryear{Brown et al.}{1982}]{Brown82}
Brown J.C., Aspin, C., Simmons J.F.L., McLean, I.S.,
1982, MNRAS, 198, 787

\bibitem[\protect\citeauthoryear{Cappa, Goss \& van der Hucht}{Cappa et al.}{2004}]{Cappa04} 
Cappa C., Goss W.M., van der Hucht K.A., 2004, AJ, 127, 2885

\bibitem[\protect\citeauthoryear{Chapman et al.}{1999}]{Chapman99}
Chapman J.M., Leitherer C., Koribalski B., Bouter R., Storey M.
1999, ApJ, 518, 890

\bibitem[\protect\citeauthoryear{Conti \& Smith}{1972}]{Conti72} 
Conti P.S., Smith L.F., 1972, ApJ, 172, 623

\bibitem[\protect\citeauthoryear{Conti \& Vacca}{1990}]{Conti90} 
Conti P.S., Vacca W.D., 1990, AJ, 100, 431

\bibitem[\protect\citeauthoryear{Conti et al.}{1983}]{Conti83} 
Conti P.S., Garmany C.D., de Loore C., Vanbeveren D., 1983, ApJ, 274, 302

\bibitem[\protect\citeauthoryear{Cousins}{1972}]{Cousins72} 
Cousins A.W.J., 1972, MNSSA, 31, 69

\bibitem[\protect\citeauthoryear{Davis et al.}{1999}]{Davis99} 
Davis J., Tango W.J., Booth A.J., ten Brummelaar T.A., Minard R.A.,
Owens S.M., 1999, MNRAS, 303, 773

\bibitem[\protect\citeauthoryear{Davis et al.}{2005}]{Davis05} 
Davis J. et al., 2005, MNRAS, 356, 1362 

\bibitem[\protect\citeauthoryear{De Marco \& Schmutz}{1999}]{DeMarco99} 
De Marco O., Schmutz W., 1999, A\&A, 345, 163

\bibitem[\protect\citeauthoryear{De Marco et al.}{2000}]{DeMarco00} 
De Marco O., Schmutz W., Crowther P.A., Hillier D.J., Dessart L.,
de Koter A., Schweickhardt J., 2000, A\&A, 358, 187

\bibitem[\protect\citeauthoryear{Dessart et al.}{2000}]{Dessart00} 
Dessart L., Crowther P.A., Hillier D.J., Willis A.J., Morris P.W.,
van der Hucht K.A., 2000, MNRAS, 315, 407

\bibitem[\protect\citeauthoryear{de Zeeuw et al.}{1999}]{deZeeuw99} 
de Zeeuw P.T., Hoogerwerf R., de Bruijne J.H.J., Brown A.G.A,
Blaauw A., 1999, AJ, 117, 354

\bibitem[\protect\citeauthoryear{Doom}{1988}]{Doom88} 
Doom C., 1988, A\&A, 192, 170

\bibitem[\protect\citeauthoryear{Eenens \& Williams}{1994}]{Eenens94} 
Eenens P.R.J., Williams P.M., 1994, MNRAS, 269, 1082

\bibitem[\protect\citeauthoryear{ESA}{1997}]{ESA97} 
ESA, 1997, The {\em Hipparcos} Catalogue, ESA SP-1200

\bibitem[\protect\citeauthoryear{Eversberg, Moffat \& Marchenko}{Eversberg et al.}{1999}]
{Eversberg99} 
Eversberg T., Moffat A.F.J., Marchenko S.V., 1999, PASP, 111, 861

\bibitem[\protect\citeauthoryear{Ganesh \& Bappu}{1967}]{Ganesh67} 
Ganesh K.S., Bappu M.K.V., 1967, Kodaikanal Obs. Bull. Ser. A, 183, 77

\bibitem[\protect\citeauthoryear{Gregory}{2005a}]{Gregory05a}
Gregory, P.C., 2005a, Bayesian Logical Data Analysis for the Physical 
Sciences: A Comparative Approach with {\em Mathematica} Support,
Cambridge University Press, Cambridge

\bibitem[\protect\citeauthoryear{Gregory}{2005b}]{Gregory05b}
Gregory, P.C., 2005b, ApJ, 631, 1198

\bibitem[\protect\citeauthoryear{Hamann \& Koesterke}{1998}]{Hamann98}
Hamann W.-R., Koesterke L., 1998, A\&A, 335, 1003

\bibitem[\protect\citeauthoryear{Hanbury Brown et al.}{1970}]{HBrown70} 
Hanbury Brown R., Davis J., Herbison-Evans D., Allen L.R., 1970,
MNRAS, 148, 103

\bibitem[\protect\citeauthoryear{Hanbury Brown, Davis \& Allen}{Hanbury Brown et al.}{1974}]{HBrown74} 
Hanbury Brown R., Davis J., Allen L.R., 1974, MNRAS, 167, 121

\bibitem[\protect\citeauthoryear{Heintz}{1978}]{Heintz78}
Heintz W.D., 1978, Double Stars, Reidel, Dordrecht

\bibitem[\protect\citeauthoryear{Henley, Stevens \& Pittard}{Henley et al.}{2005}]{Henley05} 
Henley D.B., Stevens I.R., Pittard J.M., 2005, MNRAS, 356, 1308

\bibitem[\protect\citeauthoryear{Hillier}{1989}]{Hillier89}
Hillier D.J., 1989, ApJS, 347, 392

\bibitem[\protect\citeauthoryear{Howarth \& Prinja}{1989}]{Howarth89}
Howarth I.D., Prinja R.K., 1989, ApJ, 69, 527

\bibitem[\protect\citeauthoryear{Hummel et al.}{1993}]{Hummel93}
Hummel C.A., Armstrong J.T., Quirrenbach A., Buscher D.F., Mozurkewich D., 
Simon R.S., Johnston K.J., 1993, AJ, 106, 2486

\bibitem[\protect\citeauthoryear{Ireland}{2006}]{Ireland06}
Ireland M.J., 2006, in Monnier J.D., Sch{\"o}ller M., Danchi W.C., eds, 
Proc. SPIE 6268, Advances in Stellar Interferometry

\bibitem[\protect\citeauthoryear{Johnson et al.}{1966}]{Johnson66}
Johnson H.L., Iriarte B., Mitchell R.I., Wisniewskj W.Z., 1966,
Comm. Lunar Plan. Lab 4, 99

\bibitem[\protect\citeauthoryear{Leitherer, Chapman \& Koribalski}{Leitherer et al.}{1997}]{Leitherer97}
Leitherer C., Chapman J.M., Koribalski B., 1997, ApJ, 481, 898

\bibitem[\protect\citeauthoryear{Lindegren}{1997}]{Lindegren97}
Lindegren L., 1997, ESA SP-402, 13

\bibitem[\protect\citeauthoryear{Lucy \& Abbott}{1993}]{Lucy93} 
Lucy L.B., Abbott D.C., 1993, ApJ, 405, 738

\bibitem[\protect\citeauthoryear{Malbet et al.}{2006}]{Malbet06} 
Malbet F., Petrov R.G., Weigelt G., Stee P., Tatulli E., 
Domiciano de Souza A., Millour F., 2006, in 
Monnier J.D., Sch{\"o}ller M., Danchi W.C., eds, Proc. SPIE 6268,
Advances in Stellar Interferometry

\bibitem[\protect\citeauthoryear{Martins, Schaerer \& Hillier}{Martins et al.}{2005}]{Martins05} 
Martins F., Schaerer D., Hillier D.J., 2005, A\&A, 436, 1049

\bibitem[\protect\citeauthoryear{Meynet et al.}{1994}]{Meynet94} 
Meynet G., Maeder A., Schaller G., Schaerer D., Charbonnel C., 
1994, A\&AS, 103, 97

\bibitem[\protect\citeauthoryear{Millour et al.}{2006}]{Millour06} 
Millour F. et al., preprint(astro-ph/0610936)

\bibitem[\protect\citeauthoryear{Moffat \& Marchenko}{1996}]{Moffat96} 
Moffat A.E.J., Marchenko S.V., 1996, A\&A, 305, L29

\bibitem[\protect\citeauthoryear{Moffat \& Robert}{1994}]{Moffat94} 
Moffat A.F.J., Robert C., 1994, ApJ, 421, 310

\bibitem[\protect\citeauthoryear{Moffat et al.}{1986}]{Moffat86} 
Moffat A.F.J., Vogt N., Paquin G., Lamontagne R., Barrera L.H., 
1986, AJ, 91, 1386

\bibitem[\protect\citeauthoryear{Morris et al.}{2000}]{Morris00} 
Morris P.W., van der Hucht K.A., Crowther P.A., Hillier D.J., 
Dessart L., Williams P.M., Willis A.J., 2000, A\&A, 353, 624

\bibitem[\protect\citeauthoryear{Niedzielski}{1994}]{Niedzielski94}
Niedzielski A., 1994, A\&A, 282, 529

\bibitem[\protect\citeauthoryear{Niemela \& Sahade}{1980}]{Niemela80} 
Niemela V.S., Sahade J., 1980, ApJ, 238, 244

\bibitem[\protect\citeauthoryear{Nugis \& Lamers}{2000}]{Nugis00} 
Nugis T., Lamers H.J.G.L.M., A\&A, 2000, 360, 227

\bibitem[\protect\citeauthoryear{Nugis, Crowther \& Willis}{Nugis et al.}{1998}]{Nugis98} 
Nugis T., Crowther P.A., Willis A.J., 1998, A\&A, 333, 956

\bibitem[\protect\citeauthoryear{Owocki \& Gayley}{1999}]{Owocki99} 
Owocki S.P \& Gayley K.G., 1999, in van der Hucht K.A., 
Koenigsberger G. eds, Proc. IAU 193, 
Wolf-Rayet Phenomena in Massive Stars and Starburst Galaxies

\bibitem[\protect\citeauthoryear{Pike, Stickland \& Willis}{Pike et al.}{1983}]{Pike83} 
Pike C.D., Stickland D.J., Willis A.J., 1983, The Observatory, 103, 154

\bibitem[\protect\citeauthoryear{Pozzo et al.}{2000}]{Pozzo00} 
Pozzo M., Jeffries R.D., Naylor T., Totten E.J., Harmer S., Kenyon M.,
2000, MNRAS, 313, L23

\bibitem[\protect\citeauthoryear{Press et al.}{1992}]{Press92}
Press W.H., Teukolsky S.A., Vetterling W.T., Flannery B.P., 1992,
Numerical Recipes in C -- The Art of Scientific Computing, 2nd Ed.
(Cambridge University Press)

\bibitem[\protect\citeauthoryear{Robert et al.}{1989}]{Robert89}
Robert C., Moffat A.F.J., Bastien P., Drissen L., St.-Louis N., 1989
ApJ, 347, 1034

\bibitem[\protect\citeauthoryear{Schaerer \& Maeder}{1992}]{Schaerer92} 
Schaerer D., Maeder, 1992, A\&A, 263, 129

\bibitem[\protect\citeauthoryear{Schaerer, Schmutz \& Grenon}{Schaerer et al.}{1997}]{Schaerer97} 
Schaerer D., Schmutz W., Grenon M., 1997, ApJ, 484, L153

\bibitem[\protect\citeauthoryear{Schmutz et al.}{1997}]{Schmutz97} 
Schmutz W. et al., 1997, A\&A, 328, 219

\bibitem[\protect\citeauthoryear{Schulte-Ladbeck, Eenens \& Davis}{Schulte-Ladbeck et al.}{1995}]{SLadbeck95}
Schulte-Ladbeck R.E., Eenens P.R.J., Davis K., 1995, ApJ, 454, 917

\bibitem[\protect\citeauthoryear{Setia Gunawan et al.}{2001}]{Setia01} 
Setia Gunawan D.Y.A., de Bruyn A.G., van der Hucht K.A., Williams P.M., 
2001, A\&A, 368, 484

\bibitem[\protect\citeauthoryear{Simmons \& Boyle}{1984}]{Simmons84}
Simmons, J.F.L., Boyle, C.B., 1984, A\&A, 134, 368

\bibitem[\protect\citeauthoryear{Smith}{1968a}]{Smith68a} 
Smith L.F., 1968a, MNRAS, 138, 109

\bibitem[\protect\citeauthoryear{Smith}{1968b}]{Smith68b} 
Smith L.F., 1968b, MNRAS, 140, 409

\bibitem[\protect\citeauthoryear{Smith, Shara \& Moffat}{Smith et al.}{1990}]{Smith90} 
Smith L.F., Shara M.M., Moffat A.F.J, 1990, ApJ, 358, 229

\bibitem[\protect\citeauthoryear{Smith, Meynet \& Mermilliod}{Smith et al.}{1994}]{Smith94} 
Smith L.F., Meynet G., Mermilliod J.-C., 1994, A\&A, 287, 835

\bibitem[\protect\citeauthoryear{Stevens et al.}{1996}]{Stevens96} 
Stevens I.R., Corcoran M.F., Willis A.J., Skinner S.L., Pollock A.M.T.,
Nagase F., Koyama K., 1996, MNRAS, 283, 589

\bibitem[\protect\citeauthoryear{Stickland \& Lloyd}{1990}]{Stickland90} 
Stickland D.J., Lloyd C., 1990, The Observatory, 110, 1

\bibitem[\protect\citeauthoryear{St.-Louis et al.}{1987}]{StLouis87} 
St.-Louis N, Drissen L., Moffat A.F.J., Bastien P., Tapia S.,
1987, ApJ, 322, 870

\bibitem[\protect\citeauthoryear{St.-Louis et al.}{1988}]{StLouis88}
St.-Louis N., Moffat A.F.J., Drissen L., Bastien P., Robert C., 1988
ApJ, 330, 286

\bibitem[\protect\citeauthoryear{Tango \& Davis}{2002}]{Tango02} 
Tango W.J., Davis J., MNRAS, 2002, 333, 642

\bibitem[\protect\citeauthoryear{Tango et al.}{2006}]{Tango06} 
Tango W.J. et al., 2006, MNRAS, 370, 884

\bibitem[\protect\citeauthoryear{Tokovinin et al.}{1999}]{Tokovinin99} 
Tokovinin A.A., Chalabaev A., Shatsky N.I., Beuzit J.L., 1999, A\&A, 346, 481

\bibitem[\protect\citeauthoryear{Tuthill et al.}{2004}]{Tuthill04} 	
Tuthill P.G., Davis J., Ireland M., North J.R., O'Byrne J, 
Robertson J.G., Tango W.J., 2004, Proc. SPIE, 5491, 499

\bibitem[\protect\citeauthoryear{Vacca, Garmany \& Shull}{Vacca et al.}{1996}]{Vacca96}
Vacca W.D., Garmany C.D., Shull M., 1996, ApJ, 460, 914

\bibitem[\protect\citeauthoryear{van der Hucht}{1992}]{Hucht92} 
van der Hucht K.A., 1992, A\&ARv., 4, 123

\bibitem[\protect\citeauthoryear{van der Hucht}{2001}]{Hucht01} 
van der Hucht K.A., 2001, New Astron. Rev., 45, 135

\bibitem[\protect\citeauthoryear{van der Hucht et al.}{1988}]{Hucht88} 
van der Hucht K.A., Hidayat B., Admiranto A.G., Supelli K.R., Doom C.,
1988, A\&A, 199, 217

\bibitem[\protect\citeauthoryear{van der Hucht et al.}{1996}]{Hucht96} 
van der Hucht K.A. et al., 1996, A\&A, 315, L193

\bibitem[\protect\citeauthoryear{van der Hucht et al.}{1997}]{Hucht97} 
van der Hucht K.A. et al., 1997, New Astron., 2, 245

\bibitem[\protect\citeauthoryear{van Leeuwen}{2005}]{Leeuwen05a}
van Leeuwen F., A\&A, 2005, 439, 805

\bibitem[\protect\citeauthoryear{van Leeuwen \& Fantino}{2005}]{Leeuwen05b}
van Leeuwen F., Fantino E., A\&A, 2005, 439, 791
 
\bibitem[\protect\citeauthoryear{Willis, Schild \& Stevens}{Willis et al.}{1995}]{Willis95} 
Willis A.J., Schild H., Stevens I.R., 1995, A\&A, 298, 549

\end{thebibliography}
\end{document}